\documentclass{PoS}

\usepackage{graphicx}
\usepackage{amssymb,amsmath}
\newcommand{\xv}{{\mathbf x}}
\newcommand{\tr}{\textrm{Tr}}

\title{New algorithms for finite density QCD}

\ShortTitle{New algorithms for finite density QCD}

\author{ {D\'enes Sexty}%
         \thanks{I gratefully thank my collaborators
  G.~Aarts, F.~Attanasio, L.~Bongiovanni, Sz.~Bors\'anyi, Z.~Fodor,
B.~J\"ager, S.D.~Katz, E.~Seiler and I.-O.~Stamatescu.
}\\
        ITP, University of Heidelberg\\
        Department of Physics, University of Wuppertal \\
        E-mail: \email{sexty.denes@gmail.com}}

\abstract{ Recent progress of the complex Langevin method 
 and the Lefschetz thimble in connection with the sign problem 
is reviewed. These methods rely on the complexification of the 
original field manifold
and they allow direct simulations of theories with non-real measures. 
Similarities and differences of the two approaches are pointed out.
Results using the complex Langevin method, which allows simulations 
to evade the sign problem in full QCD, are presented.
Promising results of the thimble approach for non-gauge 
theories are also discussed.
    }

\FullConference{The 32nd International Symposium on Lattice Field Theory,\\
		23-28 June, 2014\\
		Columbia University New York, NY}

\newcommand{\bea}{\begin{eqnarray}}
\newcommand{\eea}{\end{eqnarray}}

\begin{document}

\section{Introduction}

Investigations of the non-zero density phases of QCD
are hampered by the sign problem. The theory is defined with the 
path integral 
\bea
Z = \int DU e^{-S_g[U]} \textrm{Det} M(U,\mu)
\eea
with the gauge action $S_g[U]$ and the determinant of the fermion matrix 
$M(U,\mu)$.
For nonzero chemical potential $\mu$, the determinant becomes 
complex, thus importance sampling based simulation methods are not 
applicable. For small chemical potentials, the ensemble is 
close to the $\mu=0$ ensemble, thus there are various options to
evade the sign problem, such as reweighting, 
 Taylor expansion, analytic continuation from imaginary chemical 
potentials, etc. (see the reviews 
\cite{deForcrand:2010ys,Aarts:2013bla} and citations therein).
Most of these methods work by simulating in an ensemble 
with a positive measure, where naive simulations can be performed, and 
then trying to recover the original, non-positive ensemble with a 
clever choice of 
the observables, or some kind of expansion or extrapolation. 
These methods fail to recover the original ensemble if it 
is 'too far' from the positive ensemble, usually they are limited to the 
region $ \mu /T < \textrm{O}(1) $ .

In this paper I'm going to review recent progress 
in two {\it direct} simulation methods, which allow 
simulations of the non-positive ensemble we are interested in directly.
They are both based on analytical continuation: they expand the 
field manifold into the complex plane.

The first of these methods is the complex Langevin method \cite{parisi}.
It is the straightforward extension of the idea 
of stochastic quantisation \cite{parisiwu} to complex actions. After some
initial excitement in the 80's, interest in the method 
has largely disappeared, as it was noticed that the method in some 
cases gives wrong results, even though in other cases it works in spite of 
a hard sign problem. 
In recent years,
renewed interest in the method has produced studies 
with many important new results. 
Our theoretical understanding of the failures and successes of the 
method has improved substantially, with some details still missing 
(see Sec.~\ref{clesec} for details). 
Based on this understanding, an important ingredient 
for the simulations of gauge theories called gauge cooling 
has been introduced \cite{Seiler:2012wz}. This procedure removes an inherent
instability of the naive complex Langevin process of gauge theories, 
connected to the complexification of the gauge degrees of freedom.
Gauge cooling was used to solve the sign problem in HDQCD
in which the spatial fermionic hoppings are dropped \cite{Seiler:2012wz} 
(see also in 
Sec.~\ref{HDQCDsec}), and it has been extended 
also to full QCD with light quarks \cite{Sexty:2013ica}.


The complex Langevin method can of course be used for various other problems
with complex actions such as the problem of 
real time evolution \cite{realtime}, as well as gauge theories with 
a $ \Theta$-term \cite{thetaterm}.

The second method for direct simulations is the Lefschetz thimble.
The idea extends the saddle point integration to a non-perturbative 
tool which is equivalent to 
the usual path integral.  
The theory defined on the path through the saddle point 
(the 'Lefschetz thimble') is much better behaved 
from the point of view of the sign problem \cite{Cristoforetti:2012su}.
After the initial proposal the idea was quickly 
realized numerically for various models with 
promising results, see Sec.~\ref{lefschnum} for references.

This article is organized as follows: in Sec.~\ref{clesec} the complex Langevin
method is briefly described. In Sec.~\ref{lefsec} an overview of the theoretical foundation of the Lefschetz thimble is given.  
In Sec.~\ref{toysec}, both methods are applied to a simple toy model to illustrate
how they evade the sign problem.
In Sec.~\ref{lefschnum}, recent numerical results using the Lefschetz thimble
are briefly reviewed.
In Sec.~\ref{HDQCDsec}, I discuss results of the HDQCD approximation.
In Sec.~\ref{kappasec}, the $\kappa$- and $\kappa_s$-expansions are described.
In Sec.~\ref{fullsec}, results of the application of CLE to full QCD are described.
Finally Sec.~\ref{conclusionssec} concludes.

\section{Complex Langevin equation}
\label{clesec}

The Complex Langevin method is based on the analytical 
continuation of a Langevin process to 
a complexified manifold. Let's consider an 
action of one real variable $S(x)$ for demonstration. 
The real Langevin equation is written as 
$ { d x (\tau) / d \tau } = - \partial_x S(x) + \eta(\tau) $,
with the Langevin time $\tau$ and the 
noise term $\eta(\tau) $ satisfying $ \langle \eta(\tau) \rangle =0$ 
and $ \langle \eta(\tau) \eta(\tau') \rangle = 2 \delta ( \tau-\tau' ) $.
If the action is complex, the drift term  $- \partial_x S(x)$ will be complex as well,
and the field $x \rightarrow z = x+iy$ will develop an imaginary part.
We define the drift for a complex field with analytical continuation to arrive 
at the complex Langevin equation (CLE)
\bea \label{CLE}
{ d z \over d \tau } = - \left. { \partial S(z) \over \partial z } 
\right. + \eta (\tau).
\eea
The noise term can also be complexified 
but for 
practical simulations it is better to keep it real \cite{Aarts:2009uq}.
 The complex Langevin process thus enlarges the manifold of the variables 
to a non-compact group, in the case of lattice gauge theories 
the SU(N) link variables become SL(N,$\mathbb{C}$), see below.
The observables of the original theory are recovered using the principle of 
analytic continuation 
\bea \label{analytic_cont}
\langle F \rangle = 
{1 \over Z} \int dx e^{-S(x)} F(x) = 
 \int dx dy P(x,y) F(x+i y)
\eea
for the observable $F(z)$, where $P(x,y)$ is the real distribution
function of the complexified variable $ z= x+i y$ of the 
equilibrated complex Langevin process.
 Note that this prescription is equivalent to averaging the 
observable for the configurations of an equilibrated complex 
Langevin process 
\bea
 \langle F \rangle = \lim_{T \rightarrow \infty } {1\over T} \int_0^T 
d \tau F(z(\tau)),
\eea
where $z(\tau)$ is the solution of (\ref{CLE}).

The Complex Langevin method is known to deliver wrong results 
in some cases, but our understanding of this 
issue has improved recently. One can prove that the 
Complex Langevin approach delivers correct results, if some conditions 
are satisfied: the action and the observables should be holomorphic 
functions, and the distribution of the variables in the 
complexified manifold should decay sufficiently fast \cite{Aarts:2009uq}.
The failure of the method is usually the 
consequence of the violation of the second condition: slow 
decay of distributions. In the case of non-zero density QCD, the first 
of these conditions is violated, the effective action  we want 
to simulate is $S_{eff} = S_g - \textrm{ln Det}~ M $, 
where the logarithm has a branch cut. In consequence the 
drift term calculated from the action has poles where the measure 
$ e^{-S_g} \textrm{Det} M $ vanishes. The theoretical understanding 
of this issue is still limited.
The logarithm in the action can lead to problems in some cases 
\cite{Mollgaard:2013qra,greensite}, but there are also cases when 
the presence of a logarithm causes no harm, as is 
typically the case when the gauge group is parametrised with the 
help of a reduced Haar measure \cite{Aarts:2011zn,Aarts:2012ft}.
In the case of QCD, we have evidence that this issue has so far 
has not shown up: 
comparison with reweighting for small chemical 
potentials shows that CLE gives correct results \cite{fullcomp}, 
the results of full QCD are reproduced with an approach 
($\kappa$-expansion) where the action is holomorphic 
\cite{Aarts:2014bwa} (see also Sec.~\ref{kappasec}), but it remains 
to be seen whether low temperatures in particular are affected.

The Complex Langevin approach can be easily generalized to a curved 
manifold such as the SU(N) group. The discretised update 
for one variable $ U \in$ SU(N) 
reads (with straightforward generalization to a lattice system):
\bea \label{gaugeupdate}
U(\tau+\epsilon) = R(\tau) U(\tau)
\eea
where the update is defined as $R(\tau) = 
\exp ( i \lambda_a ( K_a + \eta_a \sqrt{\epsilon}) ), $ with the 
generators $ \lambda_a$,  and noise terms $\eta_a$ satisfying
$ \langle \eta_a \rangle=0 , ~ \langle \eta_a \eta_b \rangle = 2 \delta_{ab}$.
The drift term
is calculated from the action $S(U)$ as 
$ K_a= -D_a S(U) $
with the left derivative $D_a$.

The complexification of the degrees of freedom for theories 
with local symmetry enlarges also the gauge degrees of freedom. The complex 
Langevin process then tries to explore these 
non-compact gauge orbits which results in large fluctuations and slow 
decay of field distributions, and thus incorrect results.
This undesirable behavior can be countered with a procedure called 
gauge cooling \cite{Seiler:2012wz,Aarts:2013uxa}. 
The 'distance' of the field configuration 
in the enlarged SL(N,$\mathbb{C}$) field space from the original 
SU(N) manifold can be quantified by the unitarity norm:
\bea \label{unitnorm}
 \textrm{Tr} ( U^+ U -1)^2 \ge 0 . 
\eea
The gauge cooling procedure reduces this quantity using the 
gauge symmetry of the link variable $U_{\nu,x}$ in direction $ \nu$ 
at position $x$ 
\bea
U_{\nu,x} \rightarrow \Omega (x) U_{\nu,x} \Omega ( x + a_\nu ),
\eea
where $ \Omega(x) \in$ SL(N,$\mathbb{C}$) is chosen by the requirement that the average of 
the unitarity norm (\ref{unitnorm}) over the lattice should be minimal.
In practice we choose $\Omega(x)$ to be proportional to the gradient 
of the unitarity norm, and apply several gauge cooling steps after each
dynamical update of the form (\ref{gaugeupdate}). For more advanced 
gauge cooling algorithms, see Ref. \cite{Aarts:2013uxa}.

The empirical observation is that the gauge cooling procedure is 
effective as long 
as the $\beta$ parameter of the gauge action is not too small. 
The minimal usable value, $\beta_\textrm{min}$, for which 
the results of the Complex Langevin simulation agree within errors with 
other approaches (where available, such as reweighting) appears to depend 
very weakly on the lattice size and the chemical potential 
\cite{Aarts:2013nja}. 
Whether a particular $\beta$ value is usable can also be judged from 
the CLE simulation itself: for too small $\beta$ values the distribution
of the variables develops a slowly decaying tail ('skirt').
In theories with different fermionic content the actual 
value of $\beta_\textrm{min}$
might be different, but it turns out this is the usual renormalization
effect: $\beta_{min}$ corresponds to the maximal usable lattice
spacing of $ a_\textrm{max} \approx 0.1 - 0.2 ~\textrm{fm}$, roughly 
the same value independently of the numbers of flavors and their masses. 
This means that the continuum limit of the theory can be reached 
in the safe region where the method is well controlled, but there is a 
drawback that large lattices have to be used for small temperatures.

\section{Lefschetz thimble}
\label{lefsec}

We are interested in calculating averages in a theory defined 
by the partition function
\bea 
Z = \int dx M(x) = \int \exp(-S(x)) D(x) dx
\eea
where the measure $M(x)$ is typically written in the form $M(x)=\exp(-S(x)) D(x)$ 
as indicated, with the action $S(x)$ 
and $D(x)$ which can be a Haar measure (or some other Jacobian), as well 
as a fermionic determinant, and the integration goes over the real axis. 
Let's first notice that we don't necessarily have to perform
the integration in our theory for real $x$ fields.
\bea 
\int_{-\infty}^{\infty} dx M(x) O(x) = \int_{C} dz M(z) O(z)
= \int_{-\infty}^{\infty} dt J(t) M(z(t)) O(z(t))
\eea
where $C$ is some contour in the complex plane parametrized as $z(t)$, 
the deformation 
of the real axis, and $J(t)= dz/dt $ is the 
Jacobian corresponding to this variable transformation. This equality is valid
as long as the measure $M(z)$ and the observable $O(z)$ don't have
 any singularities which the integration path crosses as it is deformed.
In the theories we are interested in, this is satisfied.


Typically the measure $M(x)$ is real and 
positive, so from the point of 
view of numerical simulations
 there is no reason to deform the contour into the 
complex plane. In the cases where $M(x)$ is complex it's worth 
deforming the contour, if on the new contour 
the real part of the action is more peaked, therefore sampling 
is more efficient, and/or the imaginary part 
has less fluctuations, therefore reweighting is easier. 
Fortunately these two requirements specify 
the same curves (manifolds in higher dimensions): the Lefschetz thimbles.

The stable (instable) Lefschetz thimbles are defined as the paths 
of the steepest descent (ascent) of the real part of the action
on the complex plane, starting from a critical point, 
where $ \partial_z S(z)=0$. 
The steepest descent (ascent) of 
the real part and the constancy of the imaginary part specify the 
same curves for a holomorphic measure, as 
 can be seen by examining the Cauchy-Riemann relations.
If we choose the stable thimbles 
as the domain of the integration, the real part of the action 
is squeezed closest to the 
peak, as the path follows the steepest descent.
  The imaginary part of the action $S_I$ is constant on the thimbles, 
therefore the factor $ \exp(-iS_I) $ it can be pulled out of the integral.
The sign problem does not completely disappear, it is reintroduced 
by the factor $J(t)$, which is in general complex. 
This is called the residual sign problem, and it reappears as long as 
the deformed contour has curvature (which is equivalent 
to the model being non-Gaussian). In the case where more 
than one thimble contributes, a global sign problem also appears as one 
sums contributions from different thimbles. The residual sign 
problem is usually much milder than the original one, but its dependence on
the volume is probably still exponential, preventing the 
usage of large volumes, but allowing to draw some conclusions about 
the infinite volume theory.

\section{Lefschetz thimble and Complex Langevin solution of a toy model}
\label{toysec}

\begin{figure}
\begin{center}
\includegraphics*[width=7.1cm]{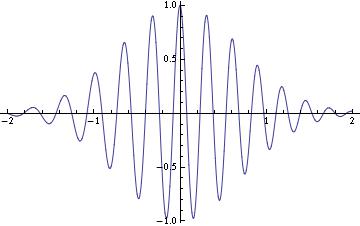}
\includegraphics*[width=7.1cm]{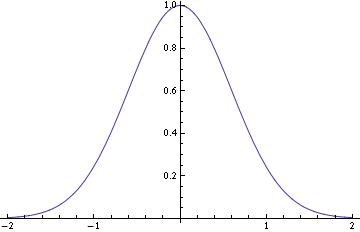}
\caption{ The measure of the Gaussian toy model defined 
in eq. ({\protect\ref{gausstoyact}})
on the real axis (left) and on the stable thimble (right) using 
$ \sigma=1+i $ and $ \lambda=20 $. }
\label{gaussian_measure}
\end{center}
\end{figure}

First a simple toy model is examined with both methods to illustrate
how they evade the sign problem. The action we are interested in is
\bea 
S(x)=\sigma x^2 + i \lambda x . 
\label{gausstoyact}
\eea
This model is Gaussian, therefore the exact solution can be calculated
analytically. Naive simulations would take the real part of the 
action, and reweight the imaginary part. Already for this simple 
model this can be 
very challenging, with lots of cancellations required 
(see left panel of Fig.~\ref{gaussian_measure}) 
to get the exact result.

\begin{figure}
\begin{center}
\includegraphics*[width=8.1cm]{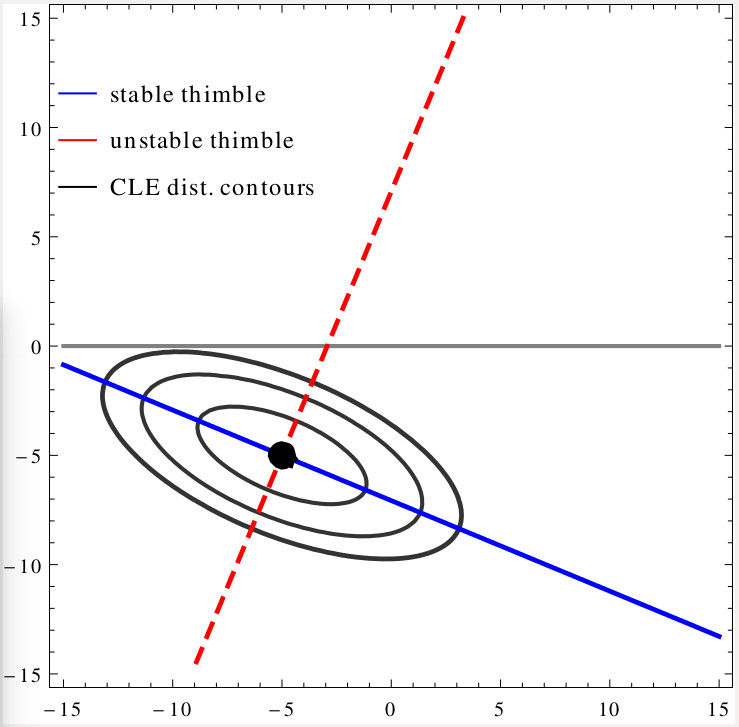}
\caption{The complexification of the Gaussian toy model defined 
in eq. ({\protect\ref{gausstoyact}}). The stable (unstable) thimble 
is given by the blue (red, dashed) line. The black ellipses signify the 
equal probability levels of the equilibrium distribution 
of the complex Langevin process. 
 }
\label{gaussian_complexified}
\end{center}
\end{figure}

This action has one critical point at $z_c=-i \lambda/2 \sigma $. 
The thimbles attached to this point are straight lines 
(see Fig.~\ref{gaussian_complexified}), thus 
the residual phase problem is absent (this simplification
happens only for non-interacting theories).
The measure on the thimble is well peaked and is easily sampled 
with a Monte-Carlo process of choice, see right panel of 
Fig.~\ref{gaussian_measure}.

In the complex Langevin approach we set up the complexified 
Langevin equation
\bea
{ d z\over d \tau } = - 2 \sigma z - i \lambda + \eta(\tau)
\eea
In this simple case the probability distribution of $z$ on the complex 
plane can be given analytically: in the equilibrium it is given 
by the Gaussian distribution, the solution of the Fokker-Planck equation
as displayed in Fig.~\ref{gaussian_complexified}. Using this distribution
in eq. (\ref{analytic_cont}) one gets the correct averages.

Note that the distribution gained with the complex Langevin process is peaked 
around the critical point of the action,
where also in the thimble approach the most dominant contributions are located.
This similarity carries over to interacting theories as well, 
first noted by Aarts \cite{Aarts:2013fpa}, 
but there are differences as well. The distribution of the complex Langevin
process is usually ``near'' to a thimble of the theory, but it is not 
necessarily 
peaked near the peak of the measure on the thimble. The 
relationship is also complicated by 
instable fixed points in the complex Langevin dynamics,
and different behavior near a zero of the measure \cite {Aarts:2014nxa}.

\section{Numerical implementations of the Lefschetz thimble}
\label{lefschnum}

In this section a short overview is given of the status 
of the implementation of the idea of the Lefschetz-thimbles in actual
numerical calculations. These implementations have several difficulties
to overcome: One has to come up with a way to sample 
the thimble with the weight $ \exp(-S_R)$, i.e. generate configurations 
that are solutions 
of the thimble equation (written for a scalar field $\phi = \phi_R + i \phi_I$)
\bea \label{steepest_desc}
{d \over d \tau} \phi_R (\tau) = - { \delta S_R [ \phi (\tau) ] \over 
\delta \phi_R } \\ \nonumber 
{d \over d \tau} \phi_I (\tau) = - { \delta S_R [ \phi (\tau) ] \over 
\delta \phi_I }
\eea
for some initial $\tau$ and critical point such that 
$ \phi(\tau=\infty) = \phi_c$,
and measure the residual phase for these configurations 
(which in general is a determinant of a large matrix, efficient calculation
methods are needed \cite{Cristoforetti:2014gsa}). The 
theory is still not free of the sign problem: the residual phase is 
than taken into account with reweighting, as it should present
a mild sign problem.
For field theories it is expected that only one critical point (corresponding 
to the global minimum of the action) needs to be 
taken into account \cite{Cristoforetti:2012su},
but in toy models with only a few degrees of freedom all critical points and their
thimbles can be important \cite{Aarts:2014nxa}. 

With the help of a mapping between a Gaussian thimble and the curved thimble 
one can perform random Metropolis sampling 
as has been demonstrated for toy models in \cite{Mukherjee:2013aga}.
One possible way to sample the thimble for a field theory is using a {\it real} 
Langevin equation \cite{Cristoforetti:2012su}
\bea
{d \over d \tau} \phi_R (\tau) = - { \delta S_R [ \phi (\tau) ] \over 
\delta \phi_R } + \eta_R \\ \nonumber 
{d \over d \tau} \phi_I (\tau) = - { \delta S_R [ \phi (\tau) ] \over 
\delta \phi_I } + \eta_I.
\eea
This equation is the same as the steepest descent equation defining the 
thimble (\ref{steepest_desc}) for zero $\eta$ noise terms. One only has 
to make sure that the noise does not drive the system off the thimble, i.e.
it has to be tangential to it. This is achieved by transporting the 
noise back to the critical point to be constrained to the 
tangential of the thimble (which is simply the space spanned by 
positive eigenvalues of the Hessian of the action at $\phi_c$), and then 
transporting it back to $\phi$. This 
procedure has been applied to the $\phi^4$ theory \cite{Cristoforetti:2013wha}
and to the Hubbard model away from half filling in 
\cite{Mukherjee:2014hsa} with promising results.

In an impressive paper \cite{Fujii:2013sra} the Hybrid Monte Carlo algorithm
was implemented for the $\phi^4$ scalar theory with a chemical
potential, where they could compare to available results using 
CLE \cite{bosegas}.  The authors studied the residual sign problem and saw
that it is indeed much milder than the original sign problem of the theory
in the naive formulation.

 Further sampling algorithms were discussed also in \cite{lefschetzlat}.

In summary the application of the Lefschetz thimble to simulations 
of field theories and toy models has yielded very encouraging results 
so far, but it is yet to be tried on a theory closer to full QCD 
(or on a theory with local symmetries).

\section{CLE treatment of the HDQCD model}
\label{HDQCDsec}

In this section the first order of a 
systematic approximation to QCD called HDQCD is described.
Consider the Wilson fermion matrix 
\bea
M(x,y)= 1 - \sum\limits_{\nu=1}^4 \kappa_\nu
\left( (1-\gamma_\nu) \exp ( {\delta_{\nu 4} \mu})
U_\nu(x) \delta_{y,x+a_\nu}  
+ (1 +\gamma_\nu) \exp ( {-\delta_{\nu 4} \mu})
U_\nu^{-1} (y) \delta_{y,x-a_\nu} 
\right),
\eea
with the hopping parameters $\kappa_\nu$ (the spatial hopping
$ \kappa_s =\kappa_1 = \kappa_2 =\kappa_3 $)
 and Euclidean Gamma-matrices $ \gamma_\nu$.
We also use the notation
\bea M = 1 - \kappa Q = 1- \kappa_s S -R, \eea
where $Q$ is the hopping matrix (it only contains terms connecting neighboring 
lattice sites), and $S (R)$ is the spatial (temporal) hopping matrix. 
The HDQCD theory corresponds to setting $ \kappa_s=0$, which can be 
justified by considering the limit 
$\kappa \rightarrow 0,\ \  \mu \rightarrow \infty, \ \ 
 \kappa e^\mu = \textrm{const.} $. This approximation can also be 
considered as the leading order (LO) of a systematic $\kappa_s$-expansion,
to be described in detail in Sec.~\ref{kappasec}.  In HDQCD,
the fermion determinant simplifies 
considerably:
\bea
 \det M_\textrm{ LO} = \prod_\xv 
 \det \left(1+ C{\cal P}_\xv \right)^2 
 \det \left(1+ C' {\cal P}_\xv^{-1}\right)^2,
 \label{e.det0} 
\eea
where $C(\mu)=(2\kappa e^\mu)^{N_\tau}$ and $C'(\mu)=C(-\mu)$, and 
$ {\cal P}_\xv$ is the Polyakov loop. The second factor 
has no contributions to the physics at large $\mu$, but it restores 
the usual symmetry of the fermion determinant 
$ \textrm{Det} M(\mu) = \textrm{Det} M(-\mu^*)^*$.

\begin{figure}
\begin{center}
\includegraphics*[width=8.1cm]{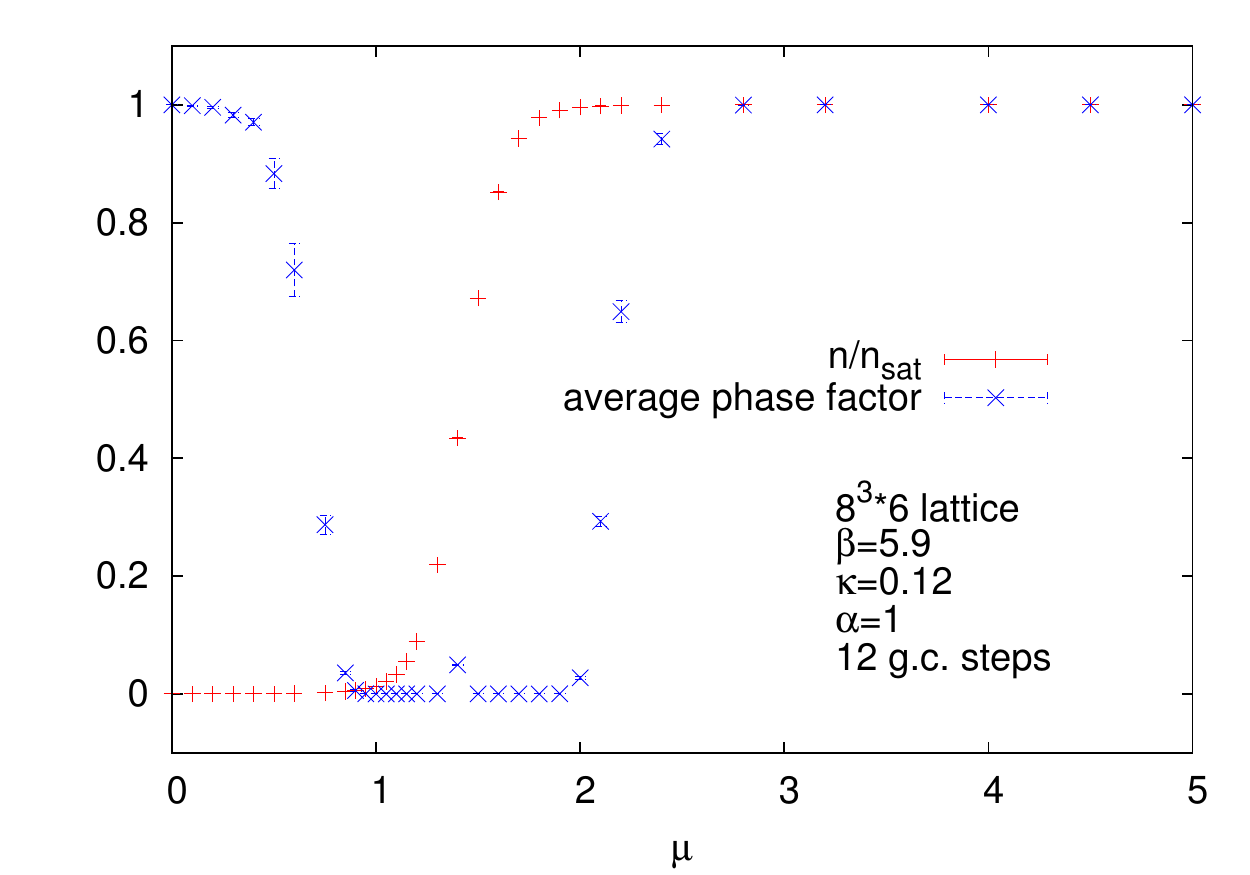}
\caption{ The density and the sign average in the HDQCD model
at a fixed temperature as a function of the chemical potential.
 }
\label{HDQCD-dens}
\end{center}
\end{figure}

This model has been investigated using the CLE and gauge cooling in Ref.
\cite{Seiler:2012wz} (for a study using reweighting, see \cite{feo}).
 In Fig.~\ref{HDQCD-dens} the behavior 
of the density $ n = (T/V )\langle \partial \textrm{ln} Z / 
\partial \mu \rangle $ (with $T$ temperature and $V$ spatial volume)
and the  average phase factor, the measure of the 
severeness of the sign problem, 
 defined as 
\bea \label{avphase}
\langle e^{2 i \varphi} \rangle =  \left\langle 
\textrm{Det} M(\mu) \over \textrm{Det} M(-\mu) \right \rangle 
\eea 
is shown.  
The average phase factor shows that for intermediate chemical
potentials the sign problem is hard, nevertheless the complex Langevin 
process is well behaved in contrast to e.g. reweighting. 

 The small chemical potential behavior resembles the low temperature 
'Silver blaze' phenomenon \cite{silverblaze}: The chemical potential 
cannot change 
the behavior of the system until it is large enough to 
create fermionic excitations.
One observes also the phenomenon of saturation: as the chemical 
potential is increased, the density of fermions can not grow past the 
value $n_{sat}$ which is realized when all states on the lattice are occupied.
Deep in the saturation state the system behaves as a pure gauge theory,
 which  can be interpreted as an 'inverse Silver Blaze' behavior:
if there is not enough thermal energy to create a hole in the fermionic sea,
the system will behave as a theory without fermions.


 Comparison with reweighting and the investigation of 
distributions shows that the gauge cooling is no longer effective 
below $ \beta_{min} \approx 5.6 $. Nevertheless one can go 
down to the confined phase by increasing the lattice size in the safe 
$\beta > \beta_{min}$ region. This allows the determination of the whole phase 
diagram of the theory on the $ \mu -T $ plane \cite{ben}.
The dropping of the spatial hoppings has the consequence that 
the phase structure of this approximation is not as rich as the phase 
structure of full QCD. 
Recently we proposed 'closing the gap' with a systematic 
expansion in the $\kappa_s$ parameter, see in Sec.~\ref{kappasec} 
and \cite{Aarts:2014bwa}.

\section{$\kappa$- and $\kappa_s$-expansion}
\label{kappasec}

As discussed in Sec.~\ref{HDQCDsec}, dropping the spatial fermionic hoppings  
simplifies the determinant such that the numerical treatment of the system
becomes cheap. Accordingly, the physics of the model becomes also simple
when compared to the richness of full-QCD. To recover some of the 
physics while keeping the numerical cost cheap, we 
propose to extend the leading order (LO) HDQCD 
simulations systematically with corrections of the 
order $\kappa_s^n$ \cite{Aarts:2014bwa}. We define two 
slightly different expansions: the $\kappa$-expansion expands the 
fermionic determinant as
\bea
\label{eq:kappaexp}
 \det M = \det(1-\kappa Q) = \exp\sum\limits_{n=1}^{\infty}  
- {\kappa^n \over n} \tr\,Q^n,
\eea
with straightforward generalization for more than one flavors. The hopping 
matrix $Q$ connects neighboring sites, therefore in the trace only even 
powers contribute. The fermionic observables are calculated similarly, using 
their definition on the expanded form, e.g. the fermionic density 
$ \langle n \rangle = (T/V) \partial \textrm{ln} Z  / \partial \mu $.

The $\kappa_s$-expansion is defined by the identity
\bea
\det M = \det (1-R) \left( 1 - \frac{1}{1-R} \kappa_s  S \right)
 = \det (1-R) \exp \sum\limits_{n=1}^{\infty}  
- \frac{\kappa_s^n}{n} \tr\left(  \frac{1}{1-R}  S \right)^n.
 \label{kappasdet}
\eea
 Since $S$ contains spatial hoppings, again only even powers contribute. 
The determinant (and the inverse) of the temporal 
part can be calculated analytically, and it gives back the 
HDQCD model in eq. (\ref{e.det0}).

These expansions can be implemented very efficiently in the complex Langevin 
simulations.  The contribution to the drift of the $\kappa$-expansion (for a single flavor) is
given by the trace
\bea
K_{x\nu a}&=&  -\sum_{n=1}^\infty \kappa^n\tr \left( Q^{n-1} 
D_{x\nu a} Q \right).
\eea

In the $\kappa_s$-expansion,  the contribution from the LO temporal 
determinant in Eq.\ (\ref{kappasdet}) given by the drift at LO \cite{Aarts:2008rr}. The second factor contributes with the traces
\bea
K_{xia} &=&  - \sum\limits_{n=1}^\infty  \kappa_s^n 
\tr \left({1\over 1-R} (D_{xia} S)  
 \left[ {1 \over 1-R}  S \right]^{n-1} \right), \nonumber \\
K_{x4a} &=&  - \sum\limits_{n=1}^\infty \kappa_s^n 
\tr \left( {1 \over 1-R } (D_{x4a} R)  
 \left[  {1 \over 1-R}  S \right]^{n} \right), \;\;\;\; 
\eea
for spatial and temporal links, correspondingly.
These traces are estimated using a noisy vector, 
i.e.\ by choosing a Gaussian random vector $ \alpha_i$  
(here $i$ represents space-time, color and Dirac indices) with 
$\langle \alpha_i \rangle=0$, $\langle \alpha^*_i \alpha_j \rangle= \delta_{ij}$, and 
then constructing the drift as 
\bea
K_{x\nu a}= \langle\alpha^* (D_{x\nu a} Q)  s\rangle, \quad\quad
  s = -\sum_n \kappa^n Q^{n-1} \alpha, \quad
\eea
for the $\kappa$-expansion.
The dominant numerical cost of including fermions, with corrections up to $\kappa^n$, is thus $n-1$ multiplications with the sparse matrix $Q$. 
This should be contrasted with the implementation of full QCD, where the 
fermionic matrix is usually inverted with a conjugate 
gradient algorithm, which requires a multiplication per iteration 
until convergence. For the $\kappa_s$-expansion the drift is 
computed similarly, with additional multiplications of 
the temporally dense but spatially diagonal matrix $(1-R)^{-1}$, as required.

\begin{figure}
\begin{center}
\includegraphics*[width=7.5cm]{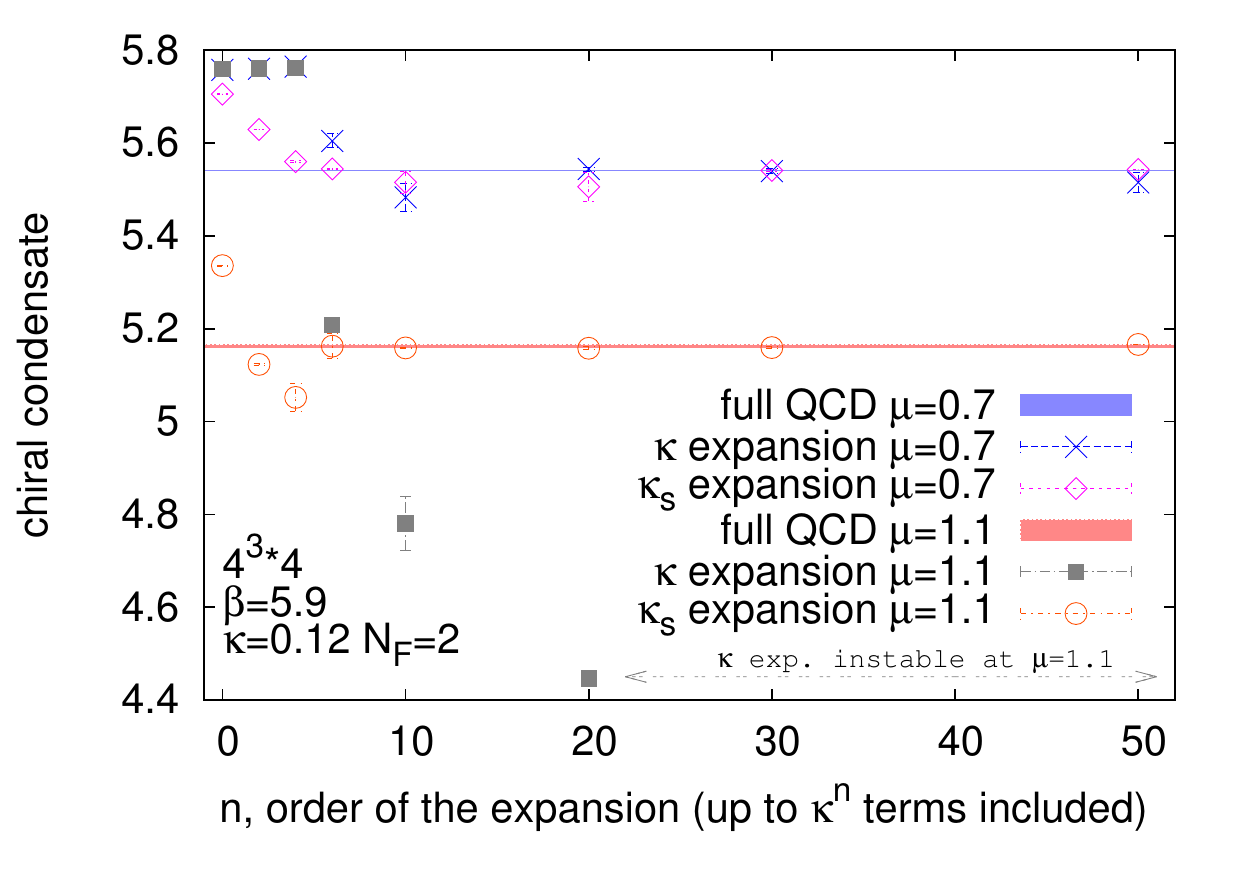}
\includegraphics*[width=7.5cm]{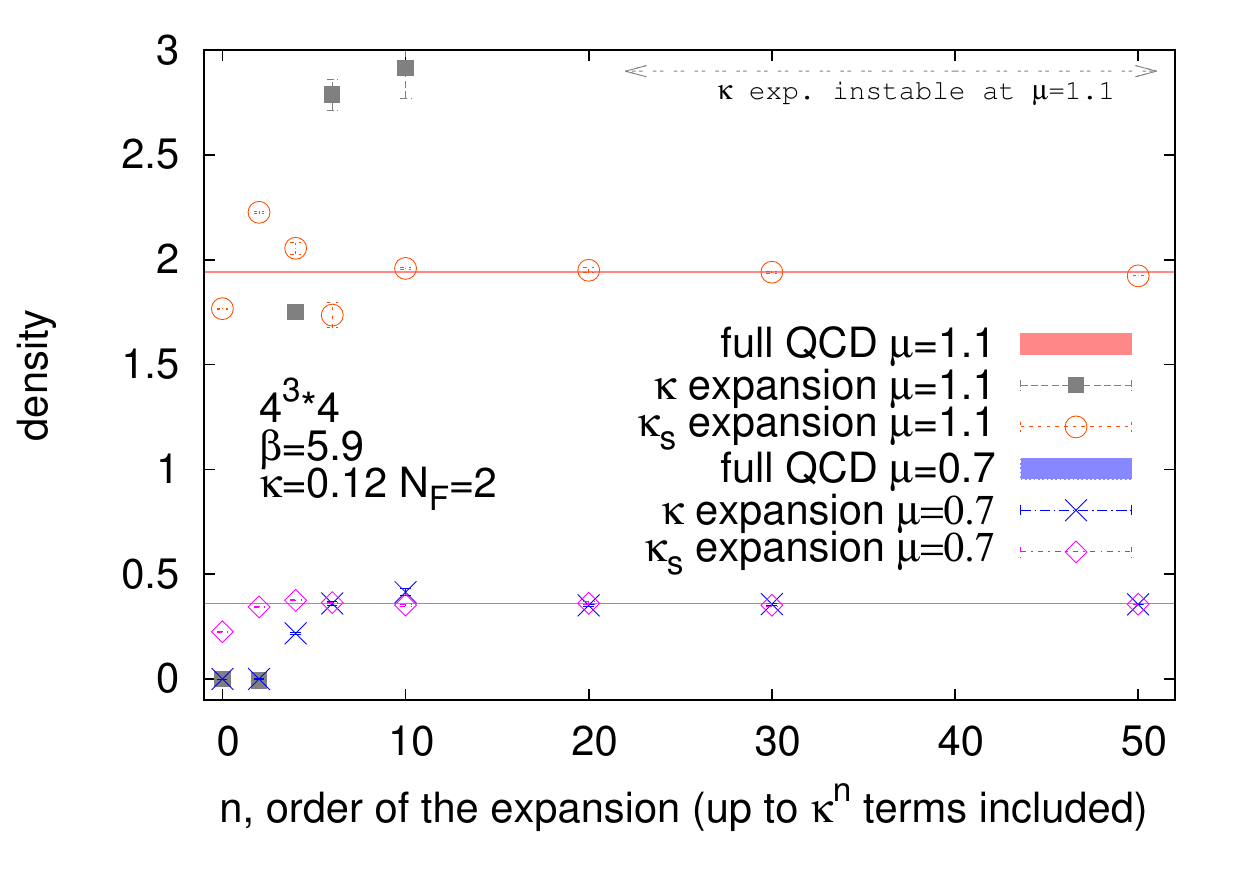}
\caption{ The results of the $\kappa-$ and $\kappa_s$-expansions
as a function of the order of corrections included and the full QCD results. 
 }
\label{kappasexp-conv}
\end{center}
\end{figure}

The convergence of the expansions defined above is illustrated on Fig.~\ref{kappasexp-conv}. Note that the procedure outlined above allows calculation
of the $\kappa$- and $\kappa_s$-expansion to very high orders, where 
convergence can be checked explicitly. Earlier studies used the 
loop expansion where higher orders become prohibitively hard, so 
only $ \kappa^2$ corrections were calculated 
\cite{Bender:1992gn,Blum:1995cb,noisymonte,Aarts:2002,feo} in connection with the full 
gauge action. Also introducing a strong coupling expansion for 
the gauge action allows 
calculations up to the $\kappa^4$ order \cite{Fromm:2011qi,Fromm:2012eb,Greensite:2013vza,Langelage:2013paa,Langelage:2014vpa}.

\begin{figure}
\begin{center}
\includegraphics*[width=11.5cm]{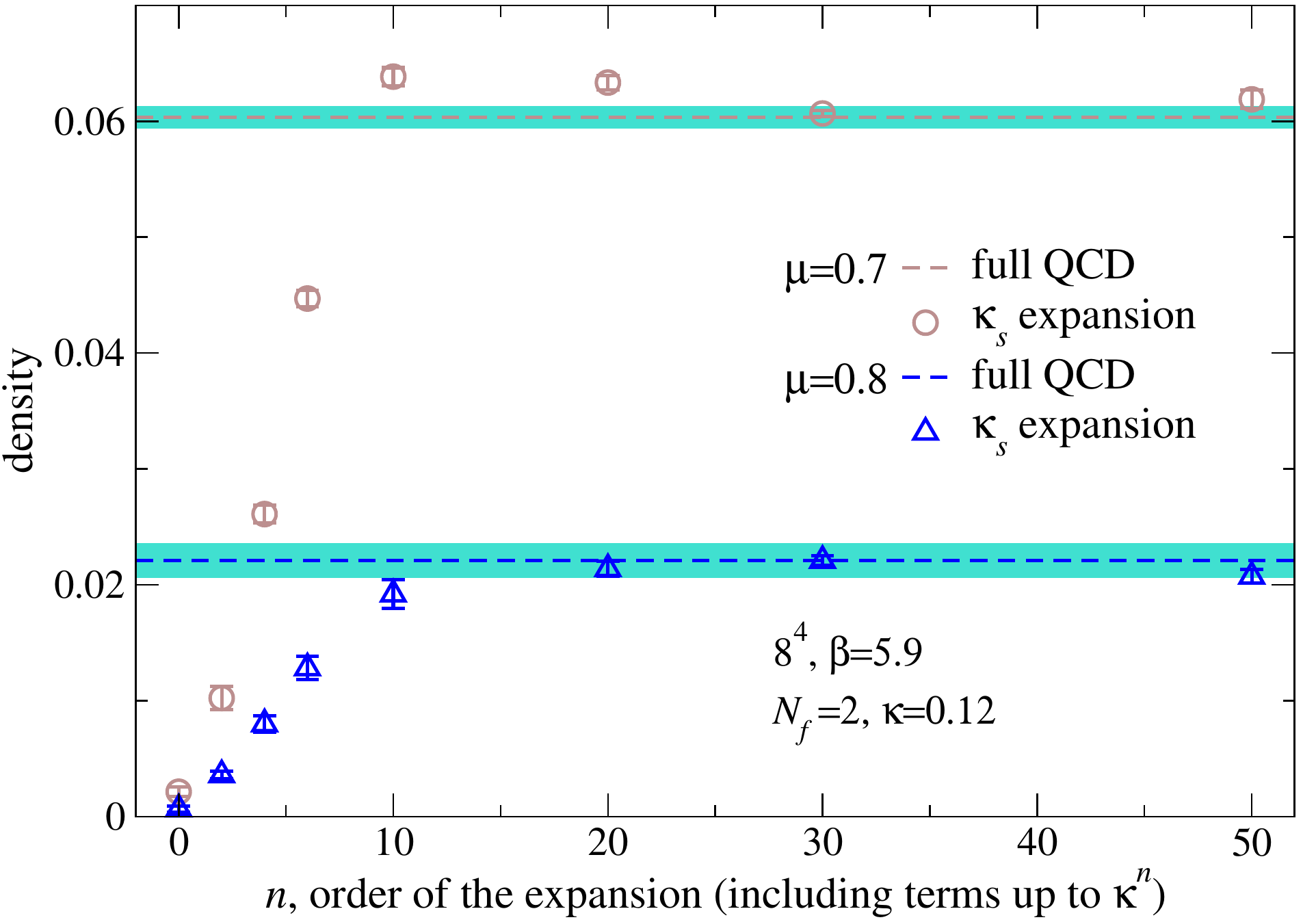}
\caption{ The density as a function of the order of corrections and the 
full QCD results on 
a $8^4$ lattice at two different chemical potentials.
 }
\label{kappasexp-biglat}
\end{center}
\end{figure}

The $\kappa$- and $\kappa_s$-expansions have different merits: the 
$\kappa$-expansion can be calculated cheaply, and is free of the problem 
of the  poles (as it is written in the exponent), but convergence 
properties at high chemical potential are not good. The $\kappa_s$-expansion
has better convergence properties at high $\mu$, but it is slightly more 
expensive, as the temporal part of the determinant is taken into account 
analytically, using a logarithm in the action. The convergence radius of the 
expansions does not seem to depend on the (temporal
 or spatial) size of the lattice,
see in Fig.~\ref{kappasexp-biglat}. In this case we tested the theory below the 
deconfinement transition, for two different $\mu$ values, and the rapid 
change in the density indicates the presence of the onset transition.

 The fact that the expansion converges (for a certain parameter range) to the 
results of full QCD validates
both theories: the expansion is used within its convergence radius and the full theory is apparently not
influenced by the non-holomorphicity of the action, as the results agree
with the $\kappa$-expansion, in which there are no logarithms, and 
thus no poles in the drift.

\section{Results for full QCD}
\label{fullsec}

The improvement in the theoretical understanding of the complex Langevin 
process and the introduction of the gauge cooling procedure has allowed for  
remarkable progress: the simulation 
of full QCD at light quark masses at high chemical 
potentials \cite{Sexty:2013ica}. 
The drift term of the CLE includes a contribution from the fermionic
determinant:
\bea
K^F_{ax\nu}=  {N_F\over 4 }\textrm Tr[ M^{-1}(\mu,U) 
  D_{ax\nu} M (\mu,U)  ],
\eea
as written for the staggered formulation. 
This contribution is estimated using Gaussian
noise vectors $\alpha$
\bea
K^F_{ax\nu}={N_F\over 4 } \alpha^+ (D_{ax\nu} M) M^{-1} \alpha,
\eea
with  $ \langle \alpha^*_x \alpha_y \rangle = \delta_{xy} $. The inverse 
of the fermion matrix is calculated using the conjugate gradient algorithm, 
and this operation represents the main numerical cost of the simulation 
for small quark masses. So far unimproved staggered and Wilson 
fermions have been implemented in the CLE.

\begin{figure}
\begin{center}
\includegraphics*[width=0.49\textwidth]{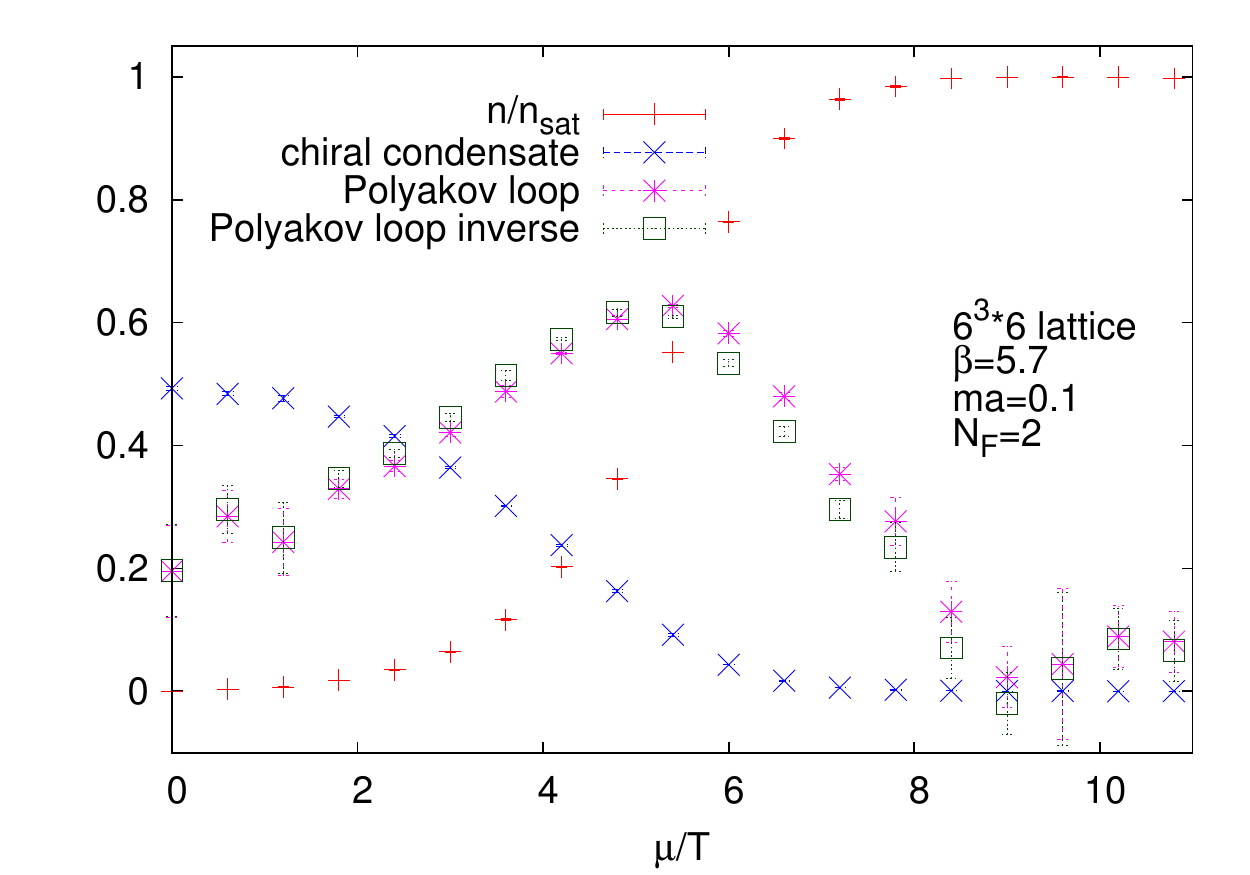}
\includegraphics*[width=0.49\textwidth]{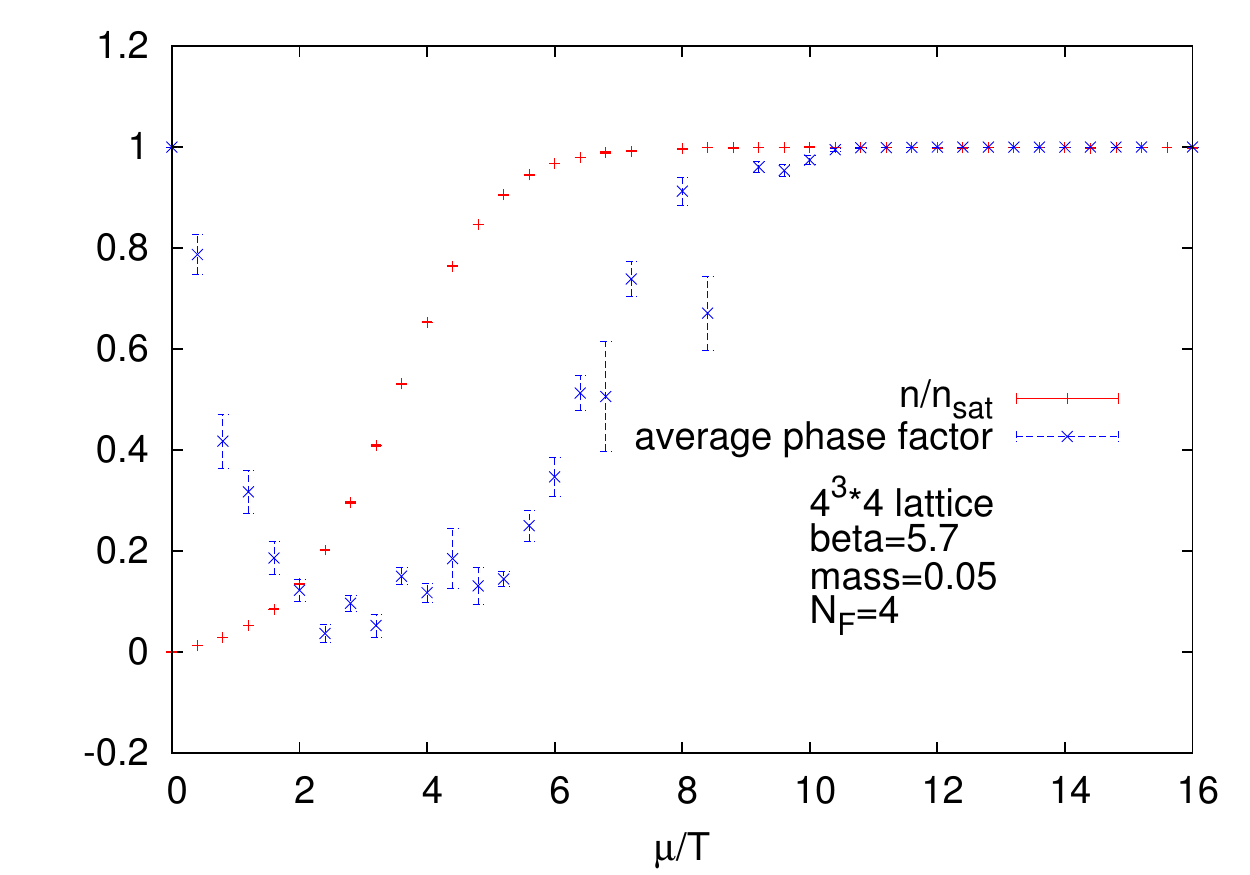}
\caption{ The fermionic density, the chiral condensate and the Polyakov loops
in full QCD with 2 flavors of staggered fermions 
as a function of the chemical potential (left panel).
 The average phase factor defined in eq. ({\protect\ref{avphase}}) as a 
function of the chemical
potential. (right panel, note that the lattice parameters are slightly different.)
 }
\label{full-horiz}
\end{center}
\end{figure}
In Fig.~\ref{full-horiz} the behavior of various observables is shown 
as a function of the chemical potential for some fixed temperature above 
the deconfinement transition. Qualitatively the behavior is the same 
as in HDQCD. One sees the fermionic density rise until saturation 
is reached. The sign problem gets hard in the dense region before saturation,
as shown in the right panel of Fig.~\ref{full-horiz}.

\begin{figure}
\begin{center}
\includegraphics*[width=0.49\textwidth]{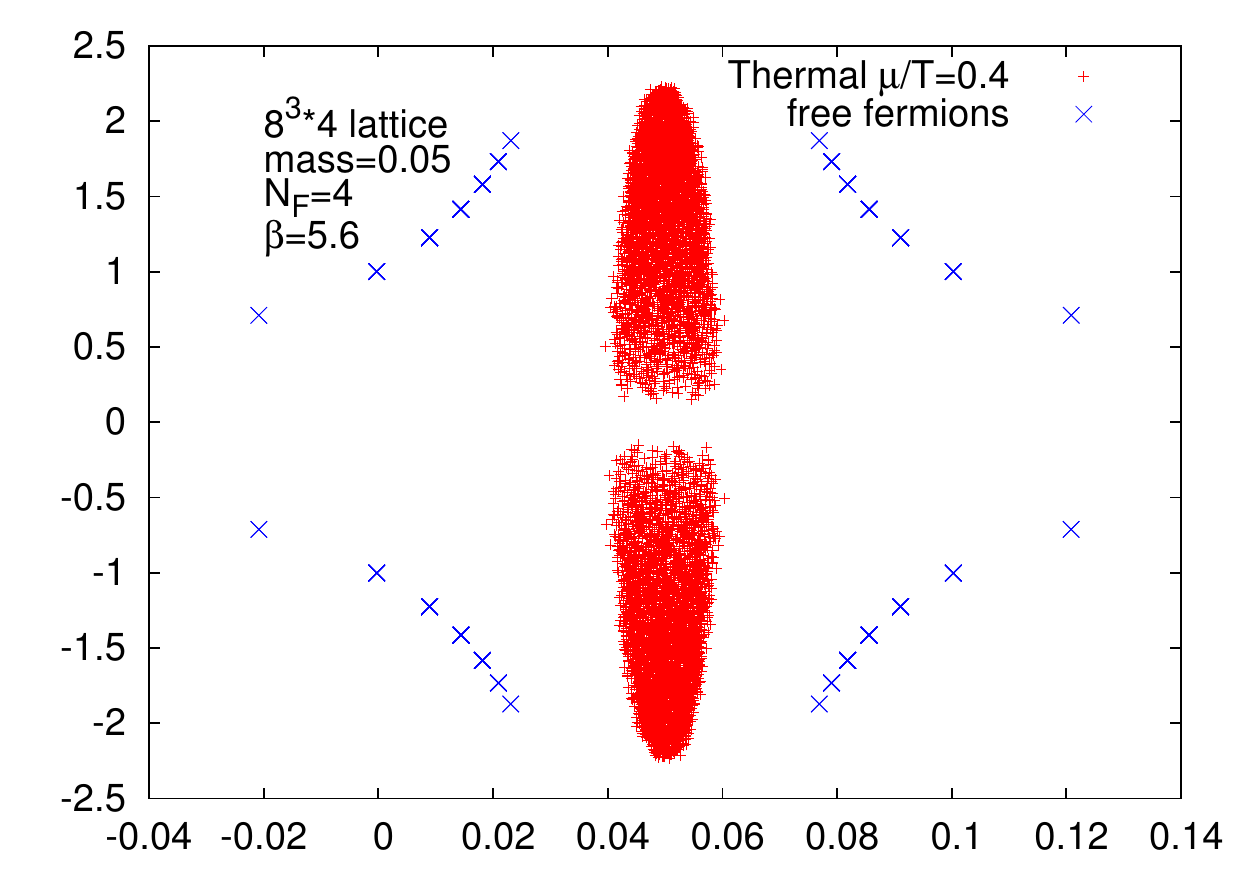}
\includegraphics*[width=0.49\textwidth]{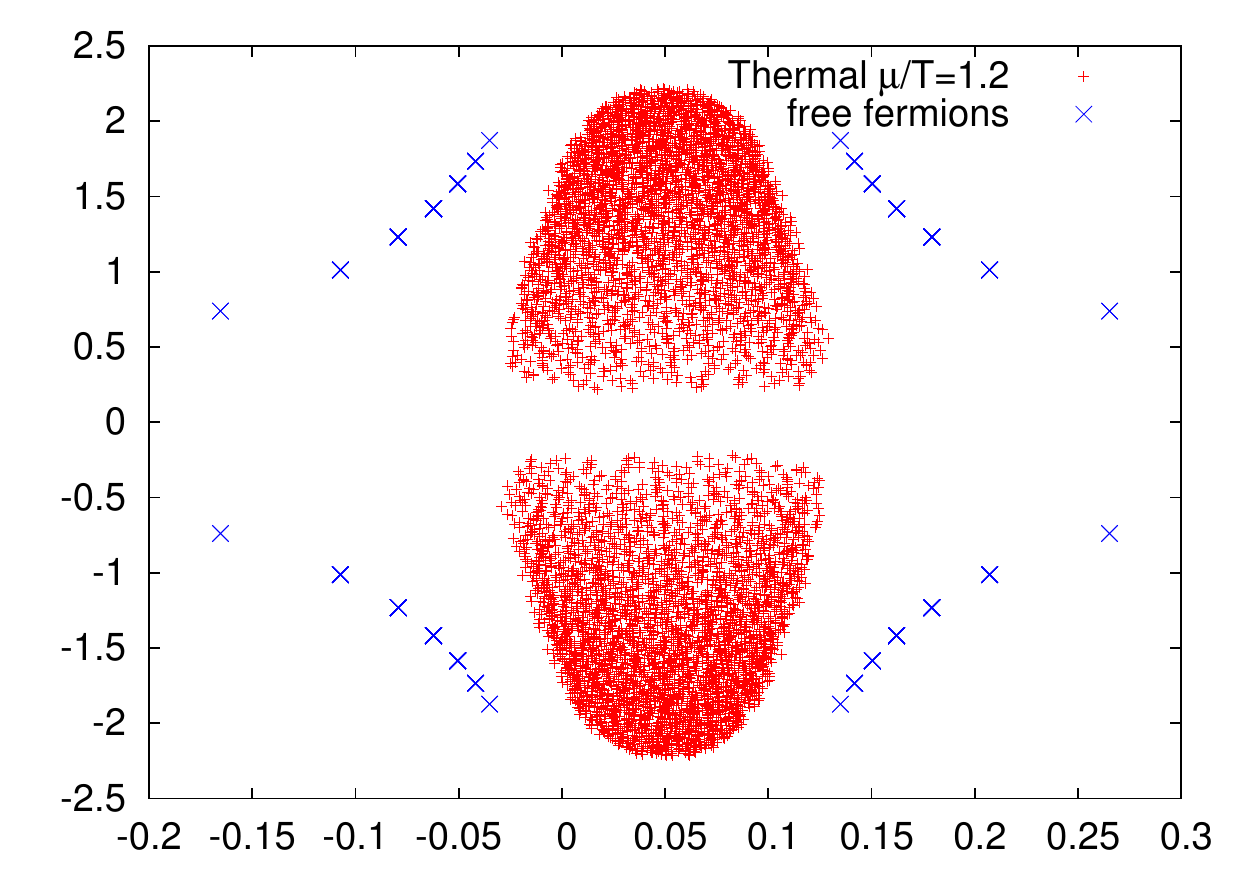}
\includegraphics*[width=0.49\textwidth]{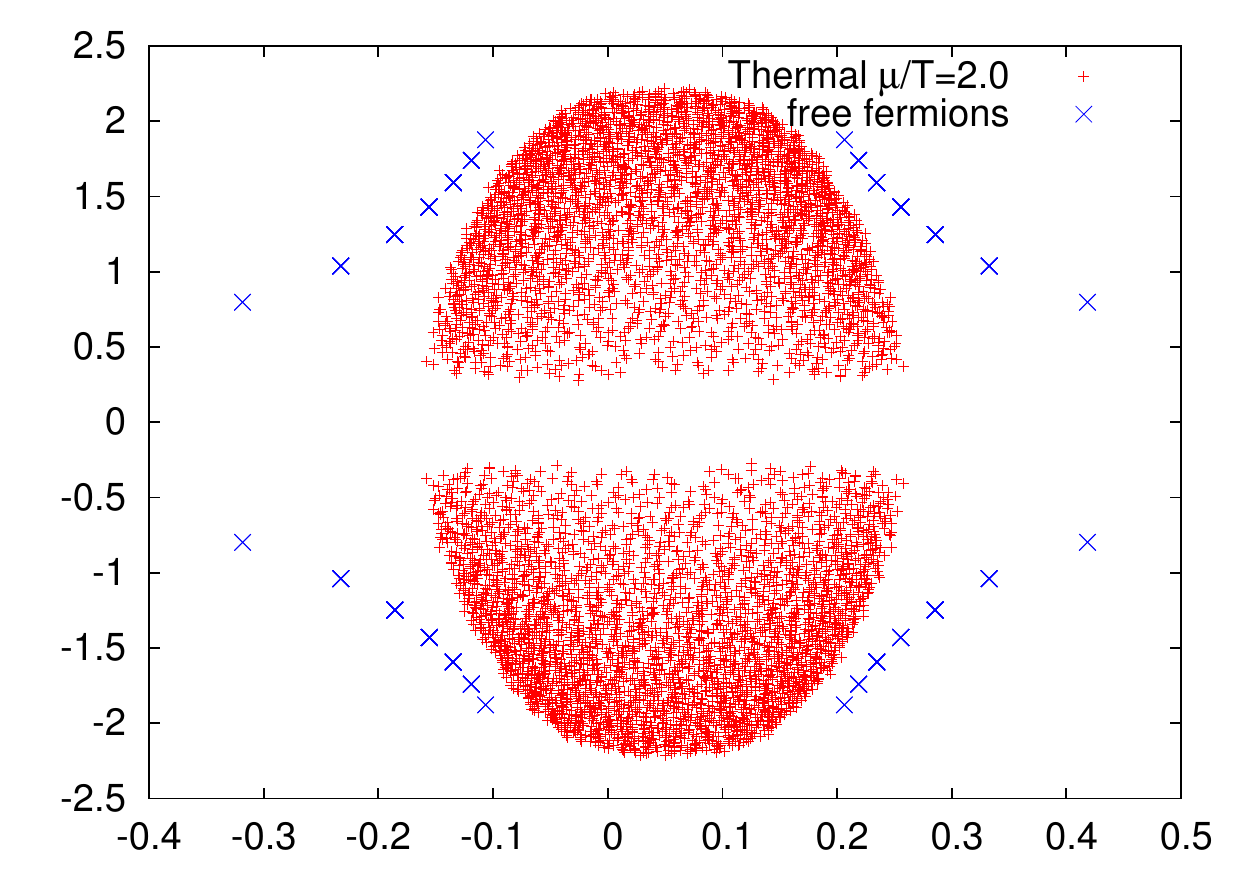}
\includegraphics*[width=0.49\textwidth]{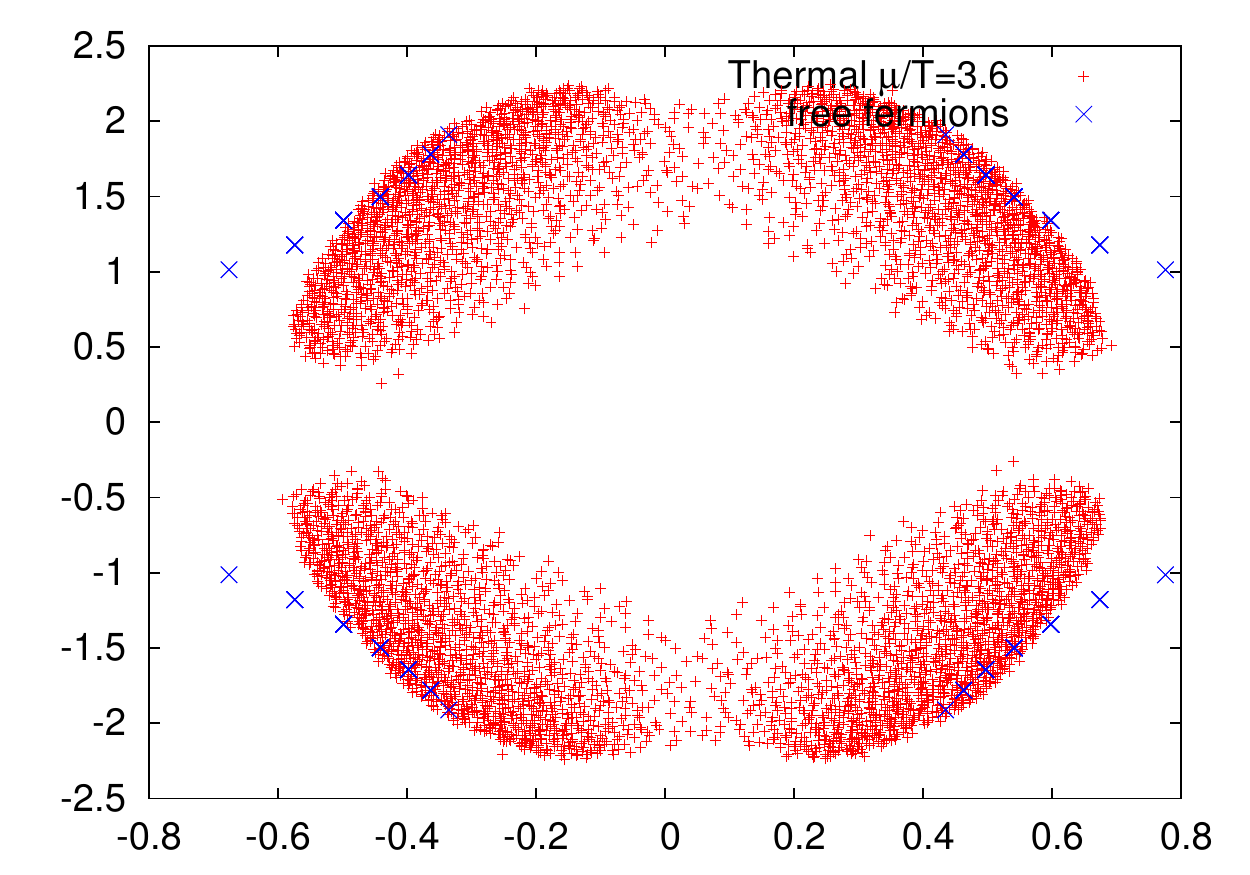}
\caption{ The spectrum of the staggered operator on the complex plane in a full 
simulation ('Thermal') as well as for a free fermionic theory
at a fixed temperature for four different chemical potentials as indicated.
 }
\label{full-spectra}
\end{center}
\end{figure}

The spectrum of the staggered operator is shown in Fig.~\ref{full-spectra}.
The temperature is fixed at some value above the deconfinement transition,
and the chemical potential is increased. For zero chemical potential
the staggered operator is antihermitian with a real shift (the mass).
At small chemical potentials the spectrum remains close to a mass shifted 
imaginary line, but the presence of $\mu$ starts to 'dilute' the spectrum 
also in real directions. As we increase the chemical potential further, 
the real part of the eigenvalues become significant. 
Nearing saturation (in the bottom right panel), the spectrum vacates the 
area around zero, as all excitations gain a large mass. 
Note that the distribution of the magnitude of the eigenvalues seems to 
have well defined lower limit, not allowing eigenvalues close to zero.
At this high temperature this can be seen as a thermal contribution to 
the masses of excitations. This in consequence means that even if there are 
poles in the drift for certain gauge field configurations, the theory does 
not go close to these dangerous areas. 
This argument (as well as the argument from the expansion,
spelled out in Sec.~\ref{kappasec}) tells us that at least at high temperatures 
the non-holomorphicity of the action causes no harm to the results.

These results show promise that one can  use the CLE to map the 
phase diagram of full QCD. Many important checks are still to be done, such 
as the reproduction of the known behavior of the crossover line at small
chemical potentials and comparison with (multi-parameter) reweighting methods
\cite{fullcomp}.

\section{Conclusions}
\label{conclusionssec}

I have reviewed two promising methods: the complex Langevin equation and the 
Lefschetz thimble, which evade the sign problem, thus
provide a way to study dense systems (among others). 
Both methods work by extending the field manifold 
to a complexified manifold (in a slightly different way), and 
thus enable direct simulations of complex 
actions. 

On the Lefschetz thimble, the dimensionality of the manifold 
is the same as the original real manifold, but the path is shifted onto 
the complex plane to such a path where the real part of the action 
is sharply peaked and the imaginary part is constant, thus the 
sign problem is absent (it is reintroduced in a mild way by the curvature 
of the manifold).  In the complex Langevin method, the dimensionality of the field manifold is doubled, and it is sampled by the complex Langevin 
equation, which does not need interpretation of the action as a probability 
measure, thus there is no sign problem.
The original theory is recovered by the expectation values.

The Lefschetz thimble approach has seen great interest in 
the past few years after its initial proposal, with many 
promising (numerical) results. 

The complex Langevin method has a 'troubled past' in contrast. It was
 noticed more than 30 years ago that the method can solve 
complex action problems, but it was also noticed a few years later that 
in some cases it produces wrong results. 
Theoretical understanding of this issue has improved a lot in the recent 
years. This better understanding lead to the introduction of the gauge-cooling
procedure which allows complex Langevin simulations of gauge theories.
This in turn allowed the simulation of full QCD with light quarks at 
high chemical potentials for the first time.

Both methods still have a number of unanswered questions, 
but in my opinion they present the most promising means of solving the sign
problem and exploring the phase diagram of full QCD for nonzero 
chemical potentials.


\begin{thebibliography}{99}


\bibitem{deForcrand:2010ys}
  P.~de Forcrand,
  PoS LAT {\bf 2009} (2009) 010
  [arXiv:1005.0539 [hep-lat]].


\bibitem{Aarts:2013bla}
  G.~Aarts,
  PoS LATTICE {\bf 2012} (2012) 017
  [arXiv:1302.3028 [hep-lat]].


\bibitem{parisi}
  G.~Parisi,
  Phys.\ Lett.\  B {\bf 131} (1983) 393.

\bibitem{parisiwu}
  G.~Parisi and Y.~s.~Wu,
  Sci.\ Sin.\  {\bf 24} (1981) 483.


\bibitem{Seiler:2012wz}
  E.~Seiler, D.~Sexty and I.~-O.~Stamatescu,
  Phys.\ Lett.\ B {\bf 723} (2013) 213
  [arXiv:1211.3709 [hep-lat]].


\bibitem{Sexty:2013ica}
  D.~Sexty,
  Phys.\ Lett.\ B {\bf 729} (2014) 108
  [arXiv:1307.7748 [hep-lat]].


\bibitem{realtime}
  J.~Berges and I.-O.~Stamatescu,
  Phys.\ Rev.\ Lett.\  {\bf 95} (2005) 202003
  [hep-lat/0508030];
  J.~Berges, S.~Borsanyi, D.~Sexty and I.-O.~Stamatescu,
  Phys.\ Rev.\ D {\bf 75} (2007) 045007
  [hep-lat/0609058];
%
  J.~Berges and D.~Sexty,
  Nucl.\ Phys.\ B {\bf 799} (2008) 306
  [arXiv:0708.0779 [hep-lat]].


\bibitem{thetaterm}
  L.~Bongiovanni, G.~Aarts, E.~Seiler, D.~Sexty and I.~O.~Stamatescu,
  arXiv:1311.1056 [hep-lat] and in prep.
; L.~Bongiovanni et. al. this proceedings.
 

 


\bibitem{Cristoforetti:2012su}
  M.~Cristoforetti {\it et al.}  [AuroraScience Collaboration],
  Phys.\ Rev.\ D {\bf 86} (2012) 074506
  [arXiv:1205.3996 [hep-lat]].


\bibitem{Aarts:2009uq}
  G.~Aarts, E.~Seiler and I.~-O.~Stamatescu,
  Phys.\ Rev.\ D {\bf 81} (2010) 054508
  [arXiv:0912.3360 [hep-lat]].


\bibitem{Mollgaard:2013qra}
  A.~Mollgaard and K.~Splittorff,
  Phys.\ Rev.\ D {\bf 88} (2013) 116007
  [arXiv:1309.4335 [hep-lat]].


\bibitem{greensite}
J.~Greensite,
  arXiv:1406.4558 [hep-lat].
 ; J.~Greensite, this proceedings


\bibitem{Aarts:2011zn}
  G.~Aarts and F.~A.~James,
  JHEP {\bf 1201} (2012) 118
  [arXiv:1112.4655 [hep-lat]].


\bibitem{Aarts:2012ft}
  G.~Aarts, F.~A.~James, J.~M.~Pawlowski, E.~Seiler, D.~Sexty and I.~O.~Stamatescu,
  JHEP {\bf 1303} (2013) 073
  [arXiv:1212.5231 [hep-lat]].


\bibitem{fullcomp}
 Z.~Fodor, S.D.~Katz, D.~Sexty, in preparation.


\bibitem{Aarts:2014bwa}
  G.~Aarts, E.~Seiler, D.~Sexty and I.~O.~Stamatescu,
  arXiv:1408.3770 [hep-lat];
  I.~O.~Stamatescu et. al., this proceedings.


\bibitem{Aarts:2013uxa}
  G.~Aarts, L.~Bongiovanni, E.~Seiler, D.~Sexty and I.~-O.~Stamatescu,
  Eur.\ Phys.\ J.\ A {\bf 49} (2013) 89
  [arXiv:1303.6425 [hep-lat]].


\bibitem{Aarts:2013nja}
  G.~Aarts, L.~Bongiovanni, E.~Seiler, D.~Sexty and I.~-O.~Stamatescu,
  arXiv:1310.7412 [hep-lat].


\bibitem{Aarts:2013fpa}
  G.~Aarts,
  Phys.\ Rev.\ D {\bf 88} (2013) 9,  094501
  [arXiv:1308.4811 [hep-lat]].


\bibitem{Aarts:2014nxa}
  G.~Aarts, L.~Bongiovanni, E.~Seiler and D.~Sexty,
  arXiv:1407.2090 [hep-lat], to appear in JHEP.

\bibitem{Cristoforetti:2014gsa}
  M.~Cristoforetti, F.~Di Renzo, G.~Eruzzi, A.~Mukherjee, C.~Schmidt, L.~Scorzato and C.~Torrero,
  Phys.\ Rev.\ D {\bf 89} (2014) 114505
  [arXiv:1403.5637 [hep-lat]].



\bibitem{Mukherjee:2013aga}
  A.~Mukherjee, M.~Cristoforetti and L.~Scorzato,
  Phys.\ Rev.\ D {\bf 88} (2013) 5,  051502
  [arXiv:1308.0233 [physics.comp-ph]].




\bibitem{Cristoforetti:2013wha}
  M.~Cristoforetti, F.~Di Renzo, A.~Mukherjee and L.~Scorzato,
  Phys.\ Rev.\ D {\bf 88} (2013) 5,  051501
  [arXiv:1303.7204 [hep-lat]].




\bibitem{Mukherjee:2014hsa} 
  A.~Mukherjee and M.~Cristoforetti,
  arXiv:1403.5680 [cond-mat.str-el].


\bibitem{Fujii:2013sra}
  H.~Fujii, D.~Honda, M.~Kato, Y.~Kikukawa, S.~Komatsu and T.~Sano,
  JHEP {\bf 1310} (2013) 147
  [arXiv:1309.4371 [hep-lat]].


\bibitem{bosegas}
  G.~Aarts,
  Phys.\ Rev.\ Lett.\  {\bf 102} (2009) 131601
  [arXiv:0810.2089 [hep-lat]].


\bibitem{lefschetzlat}
 G.~Eruzzi, this proceedings; F.~Di~Renzo, this proceedings.


\bibitem{feo}
  R.~De Pietri, A.~Feo, E.~Seiler and I.~-O.~Stamatescu,
  Phys.\ Rev.\ D {\bf 76} (2007) 114501
  [arXiv:0705.3420 [hep-lat]].
  

\bibitem{silverblaze}
   T.~D .Cohen,
  Phys.\ Rev.\ Lett.\  {\bf 91} (2003) 222001 
  [hep-ph/0307089].


\bibitem{ben} G.~Aarts, F.~Attanasio, B.~Jaeger, E.~Seiler, I.-O.~Stamatescu and  D.~Sexty, 
in preparation; B.~Jaeger et. al. this proceedings.


\bibitem{Aarts:2008rr}
  G.~Aarts and I.~-O.~Stamatescu,
  JHEP {\bf 0809} (2008) 018
  [arXiv:0807.1597 [hep-lat]].


\bibitem{Bender:1992gn}
  I.~Bender et. al. 
  Nucl.\ Phys.\ Proc.\ Suppl.\  {\bf 26} (1992) 323.


\bibitem{Blum:1995cb}
   T.~C.~Blum, J.~E.~Hetrick and D.~Toussaint,
   Phys.\ Rev.\ Lett.\  {\bf 76} (1996) 1019
   [hep-lat/9509002].

\bibitem{noisymonte}
  T.~D.~Bakeyev and P.~de Forcrand,
  Phys.\ Rev.\ D {\bf 63} (2001) 054505
  [hep-lat/0008006];


\bibitem{Aarts:2002}
  G.~Aarts, O. Kaczmarek, F. Karsch and I.~-O.~Stamatescu,
  Nucl. Phys. Proc. Suppl.\ {\bf 106} (2002) 456
  [arXiv: hep-lat0110145].




\bibitem{Fromm:2011qi}
  M.~Fromm, J.~Langelage, S.~Lottini and O.~Philipsen,
  JHEP {\bf 1201} (2012) 042
  [arXiv:1111.4953 [hep-lat]].


\bibitem{Fromm:2012eb}
  M.~Fromm, J.~Langelage, S.~Lottini, M.~Neuman and O.~Philipsen,
  Phys.\ Rev.\ Lett.\  {\bf 110} (2013) 12,  122001
  [arXiv:1207.3005 [hep-lat]].

\bibitem{Greensite:2013vza}
  J.~Greensite, J.~C.~Myers and K.~Splittorff,
  Phys.\ Rev.\ D {\bf 88} (2013) 3,  031502
  [arXiv:1306.3085 [hep-lat]].
  

\bibitem{Langelage:2013paa}
  J.~Langelage, M.~Neuman and O.~Philipsen,
  arXiv:1311.4409 [hep-lat].


\bibitem{Langelage:2014vpa}
  J.~Langelage, M.~Neuman and O.~Philipsen,
  arXiv:1403.4162 [hep-lat].





 
\end{thebibliography}
\end{document}